\documentstyle[12pt]{article}
 \def\be{\begin{eqnarray}}
 \newcommand{\bm}[1] {\mbox{\boldmath{$#1$}}}
 \def\ee{\end{eqnarray}}
\input epsfig.sty
\begin{document} 
\pagestyle{empty}
\Huge{\noindent{Istituto\\Nazionale\\Fisica\\Nucleare}}

\vspace{-3.9cm}

\Large{\rightline{Sezione di ROMA}}
\normalsize{}
\rightline{Piazzale Aldo  Moro, 2}
\rightline{I-00185 Roma, Italy}

\vspace{0.65cm}

\rightline{INFN-1230/98}
\rightline{September 1998}

\vspace{1.cm}

\begin{center}{\large\bf A Poincar\'e Covariant Current Operator
for Low and High Energy Phenomena}
\end{center}
\vskip 1em
\begin{center} F.M. Lev$^a$,  E. Pace$^b$ and G. Salm\`e$^c$ \end{center}

\noindent{$^a$\it Laboratory of Nuclear Problems, Joint Institute
for Nuclear Research, Dubna, Moscow region 141980, Russia}

\noindent{$^b$\it Dipartimento di Fisica, Universit\`a di Roma
"Tor Vergata", and Istituto Nazionale di Fisica Nucleare, Sezione
Tor Vergata, Via della Ricerca Scientifica 1, I-00133, Rome,
Italy}

\noindent{$^c$\it Istituto Nazionale di Fisica Nucleare, Sezione
di  Roma I, P.le A. Moro 2, I-00185 Rome, Italy}

\vspace{2.cm}

\begin{abstract}
Within front-form dynamics and in the Breit frame where initial and final 
three-momenta of the system are
directed along the $z$ axis, Poincar\'e covariance constrains the
current operator only through kinematical rotations
around the $z$ axis. Therefore, in this frame the current can be taken in
the one-body form. Applications to deep inelastic structure functions in an
exactly solvable model and to the deuteron magnetic form factor are presented.

\end{abstract}

\vspace{4.5cm}
\hrule width5cm
\vspace{.2cm}
\noindent{\normalsize{Proceedings of XVI European Few-Body 
Conference on "Few-Body Problems in Physics", Autrans, June 1998. 
To appear in {\bf Few-Body Systems}, Suppl. }}

\newpage
\pagestyle{plain}

\vspace{2.cm}

 The electromagnetic (em) and
weak current operators should properly commute with the 
Poincar\'e generators and satisfy Hermiticity. The em current 
should also satisfy parity and time reversal covariance, as well as continuity
equation. For instance,  the
tensor 
$\langle P',\chi'|
J^{\mu}(x)J^{\nu}(0) |P',\chi'\rangle$ in deep inelastic scattering
will have correct transformation properties relative to the Poincar\'e group
only if both the nucleon state $|P',\chi'\rangle$, with four-momentum $P'$ and
internal wave function $\chi'$, and the operator $J^{\mu}(x)$ have  correct
transformation properties with respect to the {\em same} representation of the
Poincar\'e group.

  In ref. \cite{LPS} we investigated  within the front-form dynamics \cite{Dir}
the constraints imposed on the current operator $J^{\mu}(x)$ by extended
Poincar\'e covariance (continuous + discrete transformations), Hermiticity and
current conservation using a spectral decomposition of the current operator 
and we showed that Poincar\'e covariance of $J^{\mu}(x)$ can
take place if $J^{\mu}(0)$ satisfies Lorentz covariance \cite{LPS}.

 Let ${\cal H}$ be the space of states for an  $N$ 
particle system  and let $\Pi_i$ be
the projector onto the subspace ${\cal H}_i\equiv \Pi_i{\cal H}$
corresponding to the mass  $M_i$ and the spin  $S_i$.
Since
$J^{\mu}(0)=\sum_{ij}\Pi_i J^{\mu}(0)\Pi_j$
the operator $J^{\mu}(0)$ is fully defined by
the operators $J^{\mu}(P_i,P_j')$, acting in the internal space  and
corresponding to definite values of the masses: 
$J^{\mu}(P_i;P_j')\equiv\langle {\vec {P}}_{\bot},P^+|\Pi_iJ^{\mu}(0)\Pi_j|
{\vec {P}}_{\bot}',P^{'+}\rangle$,
with ${\vec P}_{\bot}\equiv (P_x,P_y)$ and 
$P^{\pm}=(P^0 \pm P^z)/\sqrt{2}$). In the front form  rotations around 
the $z$ axis are kinematical, while the ones around $x$ and $y$ axes are
dynamical. To take advantage of this fact, we use the Breit frame with 
the
initial and final three-momenta of the system  directed along the $z$ axis.
 In order to satisfy Poincar\'e covariance, the operator
$j^{\nu}(K{\vec e}_z;M_i,M_j)$ (which is  the current operator
$J^{\mu}(K_i,K_j')$ in this particular Breit frame,  with $K{\vec e}_z = {\vec
K}_i$) has to be covariant with respect to rotations
around the $z$ axis \cite{LPS}
\begin{eqnarray}
j^{\mu}(K{\vec e}_z;M_i,M_j)=L(u_z)^{\mu}_{\nu} D^{S_{i}}(u_z)
j^{\nu}(K{\vec e}_z;M_i,M_j)D^{S_{j}}(u_z)^{-1}
\label{56}
\end{eqnarray}
where $D^{S}(u)$ is the rank $S$ representation of the rotation $u_z$
around the $z$ axis.

Therefore for a non-interacting system the continuous Lorentz transformations 
constrain the
current $j^{\mu}(K{\vec e}_z;M_i,M_j)$  in the same
way as in the interacting case.
The same property holds for the covariance with respect to the reflection of
the $y$ axis, ${\cal P}_y$, and to the product of parity and time
reversal, $\theta$, which leave the light cone $x^+=0$ invariant and are
kinematical. Hence the constraints imposed on the  current for an
interacting system by extended Lorentz covariance can be fulfilled by a current
composed in our
frame by the sum of one-body  currents (i.e., by $J_{free}^{\mu}(0) =
\sum_{n=1}^{N} j_{free,n}^{\mu}$, where N is the number of constituents).

The Hermiticity, $j^{\mu}({\vec K};M_i,M_j)^*=
j^{\mu}({-\vec K};M_j,M_i)$, is satisfied if
\begin{eqnarray}
j^{\mu}(K{\vec e}_z;M_i,M_j)^*= 
L[r_x(\pi)]^{\mu}_{\nu} D^{S_j}[r_x(\pi)]
j^{\nu}(K{\vec e}_z;M_j,M_i) D^{S_i}[r_x(\pi)])^{-1}
\label{60}
\end{eqnarray}

\noindent where the symbol $^*$ means Hermitian conjugation in internal space
and $r_x(\pi)$ represents a rotation by $\pi$ around the $x$ axis. Equation
(\ref{60}) is a non trivial constraint  when $M_i=M_j$ (i.e., for  elastic form
factors), because in this case the rhs and the lhs contain the same operator
\cite{LPS}.

 As a first application, we will study the deep inelastic scattering in an
exactly solvable model. Because of Hermiticity,  the
hadronic tensor for a
system of mass $m$, ground state $\chi_o$ and initial momentum $P$, in the
 Breit reference frame where
${\bm P}_{\bot}={\bm q}_{\bot}=0$, $P_{z}=-P'_{z}=K>0$, can be written as 
follows
 \begin{eqnarray}
&& W^{\mu\nu}=\frac{1}{4\pi}\sum (2\pi)^4\delta^{(4)} (P+q-P')
\langle \chi_o|j^{\mu}(K\bm{e}_z;m,M') |\chi'\rangle \cdot\nonumber\\
&& \overline{\langle \chi_o|j^{\nu}(K\bm{e}_z;m,M') |\chi'\rangle}
\label{92}
\end{eqnarray}

\noindent where the sum is taken over all final states 
$|P',\chi'\rangle$ of mass $M'$.

As we have seen, the free current fulfills the extended Lorentz
covariance in our Breit frame. Furthermore in the calculation of the
structure functions only three components of the current are needed and can be
chosen unconstrained by the current conservation, while the fourth
component can be determined through the continuity equation. Therefore the
structure functions can be calculated by using the $+$ and ${\perp}$ components
of the free current  in our Breit frame, even in the case
where the final state interaction is present (cf. \cite{LPS}, \cite{LPS1}).

 Let us consider a system of two particles, each one of mass $m_o$ and spin
$(1/2, \sigma_n)$, interacting through a relativistic harmonic oscillator
potential, with ground state $\chi_o({\bm k}_{\bot},\xi,\sigma_1,\sigma_2)$
($\xi=p^+/P^+$), and mass eigenvalues $M_n = 2 \{ m_o^2 + a^2 [2(n_x+n_y+n_z)+3]
\} ^{1/2}$ \cite{LPS1}.
If the exact harmonic oscillator eigenstates are used, the
hadronic tensor and then the structure functions become sums of $\delta$ 
functions. In order to obtain
continuous functions, one can consider average values,
$\overline{F}_{1(2)}(x,Q)$, of the structure functions over small intervals of
$x$, which resemble the finite experimental resolution. Taking exactly
into account the interaction, both in the initial and in the final states, in
the Bjorken limit ($Q^2 \rightarrow \infty$, $x=Q^2/(2Pq)$) the averaged 
structure functions yield exactly the parton
model results \cite{LPS1}:
\begin{eqnarray}
&& x=\xi, \quad \overline{F}_1(x) = \overline{W}^{11} = 
\sum_{\sigma_1\sigma_2}\int |\chi_o({\bm k}_{\bot},x,\sigma_1,\sigma_2)|^2 
d{\bm k}_{\bot}/ \{ 4(2\pi)^3x(1-x) \}
\nonumber\\ 
&& W^{+\nu}= 0 , \quad  \overline{F}_2(x,Q) = 2x \overline{F}_1(x,Q).
\label{292} 
\end{eqnarray}

\indent As a second application, we calculate elastic  deuteron
form factors. Let $\Pi$ be the projector onto the subspace of bound states
$|m,S,S_z \rangle$.
 The following current:  
\begin{eqnarray} 
j^{\mu}(K\bm{e}_z;m,m)=\frac{1}{2}\{{\cal{J}}^{\mu}
(K\bm{e}_z;m,m) + {\cal{J}}^{\mu}(-K\bm{e}_z;m,m)^* \} \nonumber\\ 
  {\cal{J}}^{\mu}(-K\bm{e}_z;m,m) =
L[r_x(-\pi)]^{\mu}_{\nu}exp(\imath \pi S_x)
{\cal{J}}^{\nu} (K\bm{e}_z;m,m)
exp(-\imath \pi S_x).
\label{95'}
\end{eqnarray}
(with ${\cal{J}}^{\mu}(K\bm{e}_z;m,m) = 
\langle 0,P^+|\Pi J_{free}^{\mu}(0)\Pi | 0,P'^{+}\rangle$) is compatible with 
extended Lorentz covariance
and Hermiticity, Eq. (\ref{60}), and fulfills  current conservation \cite{LPS}.

In the elastic case one has only
$2S+1$ independent matrix elements for the em current defined in Eq.
(\ref{95'}), corresponding to the $2S+1$ elastic form factors \cite{LPS}.
Therefore, the extraction of em form factors is no more
plagued by the ambiguities which are present when
the reference frame $q^+=0$ is used (see, e.g., \cite{GrKo}). 
Only three matrix elements
 $\langle m_d S S_z| j^{\mu}(K\bm{e}_z;m_d,m_d) | m_d S S_z' \rangle$
are independent for the deuteron (e.g., $\langle m_d 1 0| j^+ | m_d 1 0
\rangle$, $\langle m_d 1 1| j^+ | m_d 1 1 \rangle$, 
$\langle m_d 1 1| j_x | m_d 1 0 \rangle$), corresponding to
the three em elastic form factors.

 In Fig. 1 we report our result for the deuteron magnetic form
factor 
\begin{equation}
  B(Q^2) = 2  \langle m_d 1 1| j_x(K\bm{e}_z;m_d,m_d) | m_d 1 0 \rangle  ^2
/(3 m_d^2)
\label{98'}
\end{equation}
corresponding to different $N-N$ interactions and the
Gari-Kr\"{u}mpelmann nucleon form factors. A reasonable agreement with the
available experimental data is achieved. A more stringent comparison with
the new TJNAF data will be possible in the near future. In Table 1 our results 
for the deuteron magnetic moment
\begin{equation}
  \mu _d = \lim_{Q \rightarrow 0}  2^{1/2} m_p \langle m_d 1 1|
j_x(K\bm{e}_z;m_d,m_d) | m_d 1 0 \rangle 
 /(Q m_d)
\label{99'}
\end{equation}
are compared with the results of Ref. 
\cite{CCKP}, obtained using the free current in the $q^+=0$ reference frame.
While the $q^+=0$ approach points to a low $P_D$, our covariant approach 
prefers
higher $P_D$ values.

Calculations for charge and quadrupole form
factors are in progress.
\begin{figure}

\epsfig{bbllx=-10mm,bblly=206mm,bburx=0mm,bbury=284mm,file=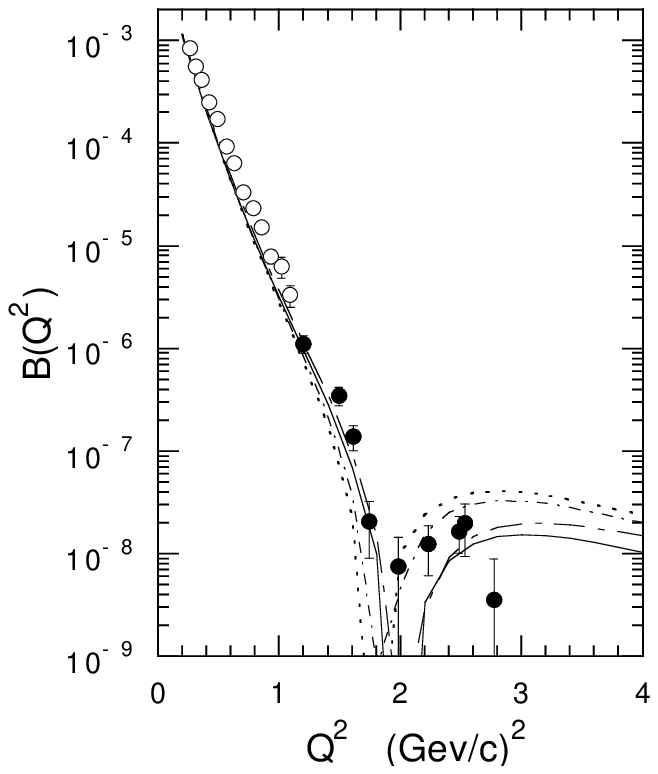}

{\bf Figure 1.} - The deuteron magnetic form factor $B(Q^2)$ obtained with the 
Gari-Kr\"{u}mpelmann nucleon form factors and different $N-N$ interactions: RSC
(dotted line), Av14 (solid line), Av18 (dashed line), Paris (dot-dashed line). 
Experimental data are from Refs. [6a] (open dots) and  [6b] (full
dots). 
\end{figure}

\begin{table}
{\bf Table 1.}{ - $\mu ^{th}_d$ for different $D$-state percentages, $P_D$
 ($\mu ^{exp}_d = 0.8574$). }
\begin{center}
\begin{tabular}{|c|c|c|c|}
\hline
Interaction & $P_D$  & $\mu _d$ (ref.\cite{CCKP}) & $\mu _d$ (this paper) \\
\hline
RSC         & 6.47   & 0.8500          & 0.8611  \\
Av14        & 6.08   & 0.8516          & 0.8608  \\
Paris       & 5.77   & 0.8531          & 0.8632  \\
Av18        & 5.76   &                 & 0.8635   \\
\hline
\end{tabular}
\end{center}
\end{table}

\end{document}